\documentclass{aa}
\usepackage[varg]{txfonts}
 
\begin{document}
 
\title{Physical characterisation of near-Earth asteroid (1620) Geographos}
\subtitle{Reconciling radar and thermal-infrared observations}

\titlerunning{Physical characterisation of (1620) Geographos}

\author{B. Rozitis\inst{1,2}
	\and S. F. Green\inst{2}
} 

\authorrunning{B. Rozitis \& S. F. Green}

\institute{Department of Earth and Planetary Sciences, University of Tennessee, Knoxville, TN 37996-1410, US
	\\\email{brozitis@utk.edu}
	\and Planetary and Space Sciences, Department of Physical Sciences, The Open University, Walton Hall, Milton Keynes, MK7 6AA, UK
} 

\date{Received XXX / Accepted XXX}

\abstract
{The Yarkovsky (orbital drift) and YORP (spin state change) effects play important roles in the dynamical and physical evolution of asteroids. Thermophysical modelling of these observed effects, and of thermal-infrared observations, allows a detailed physical characterisation of an individual asteroid to be performed.} 
{We perform a detailed physical characterisation of near-Earth asteroid (1620) Geographos, a potential meteor stream source and former spacecraft target, using the same techniques as previously used in Rozitis et al. (2013) for (1862) Apollo.} 
{We use the advanced thermophysical model (ATPM) on published light-curve, radar, and thermal-infrared observations to constrain the thermophysical properties of Geographos. The derived properties are used to make detailed predictions of the Yarkovsky orbital drift and YORP rotational acceleration, which are then compared against published measurements to determine Geographos's bulk density.}
{We find that Geographos has a thermal inertia of 340 $_{-100}^{+140}$ J m$^{-2}$ K$^{-1}$ s$^{-1/2}$, a roughness fraction of $\geq$ 50\%, and a bulk density of 2100 $_{-450}^{+550}$ kg m$^{-3}$ when using the light-curve-derived shape model with the radar-derived maximum equatorial diameter of 5.04 $\pm$ 0.07 km. It is also found that the radar observations had overestimated the z-axis in Geographos's shape model because of their near-equatorial view. This results in a poor fit to the thermal-infrared observations if its effective diameter is kept fixed in the model fitting.}
{The thermal inertia derived for Geographos is slightly higher than the typical values for a near-Earth asteroid of its size, and its derived bulk density suggests a rubble-pile interior structure. Large uncertainties in shape model z-axes are likely to explain why radar and thermal-infrared observations sometimes give inconsistent diameter determinations for other asteroids.}

\maketitle

\section{Introduction}

The asymmetric reflection and thermal re-radiation of sunlight from an asteroid's surface imposes a net force (Yarkovsky effect) and torque (Yarkovsky-O'Keefe-Radzievskii-Paddack or YORP effect). The Yarkovsky effect results in a drift in the semimajor axis of an asteroid's orbit, and the YORP effect changes its rotation period and the direction of its spin axis. Both effects are of fundamental importance for the dynamical and physical evolution of small asteroids in the solar system (see review by Bottke et al. 2006 and introduction of Rozitis et al. 2013). Furthermore, asteroid bulk densities can be determined from model-to-measurement comparisons of the Yarkovsky semimajor axis drift. The Yarkovsky effect has been detected by sensitive radar ranging for (6489) Golevka and (101955) Bennu (Chesley et al. 2003, 2014), and by deviations from predicted ephemerides over a long time span for several tens of other near-Earth asteroids including (1620) Geographos (Vokrouhlický et al. 2008; Chesley et al. 2008; Nugent et al. 2012; Farnocchia et al. 2013).  The YORP effect has been detected through observations of phase shifts in photometric light-curves of five near-Earth asteroids, which include (54509) YORP (Lowry et al. 2007; Taylor et al. 2007), (1862) Apollo (Kaasalainen et al. 2007; Ďurech et al. 2008a), (1620) Geographos (Ďurech et al. 2008b), (3103) Eger (Ďurech et al. 2012a), and (25143) Itokawa (Lowry et al. 2014). Accurate predictions of the Yarkovsky and YORP effects must take into account various thermophysical properties, which include the asteroid's size and shape, mass and moment of inertia, surface thermal properties, rotation state, and its orbit about the Sun. Recently, Rozitis et al. (2013) have produced a unified model which can simultaneously match both observed effects for (1862) Apollo using a single set of thermophysical properties derived from ground-based observations.

The object (1620) Geographos (hereafter referred to as just Geographos) is an Apollo and S-type near-Earth asteroid (Bus \& Binzel 2002), and has detections of both Yarkovsky orbital drift and YORP rotational acceleration. Its semimajor axis was found to be decreasing at a rate of 27.4 $\pm$ 5.7 m yr$^{-1}$ (mean value from Chesley et al. 2008; Nugent et al. 2012; Farnocchia et al. 2013), and its rotation rate was found to be increasing at a rate of (1.5 $\pm$ 0.2) $\times10^{-3}$ rad yr$^{-2}$ (Ďurech et al. 2008b). Photometric investigations also revealed a retrograde rotation and a very high amplitude light-curve ($\sim$2 magnitudes) indicative of a highly elongated shape (Dunlap 1974; Michalowski et al. 1994; Kwiatkowski 1995; Magnusson et al. 1996). The highly elongated shape was confirmed by radar studies, which also find it to have tapered ends and for it to contain many ridges and concavities (Ostro et al. 1995, 1996; Hudson \& Ostro 1999). Unfortunately, the radar-derived shape model does contain a north-south ambiguity because of the near-equatorial view when the radar observations were taken. Nevertheless, this unusual shape with high elongation is suggestive of a rubble-pile asteroid that was tidally distorted during a close planetary encounter (Solem \& Hills 1996; Bottke et al. 1999). Geographos's shape and rotation period ($\sim$5.2 hours) make it possible for loose material to be lofted away from the surface during close encounters with the Earth (Ryabova 2002a,b). It is not clear whether any meteors originating from Geographos using this low-velocity ejection mechanism have been detected, but two meteor streams consistent with high-velocity ejection (up to 1 km s$^{-1}$) have been identified in meteor catalogues (Ryabova 2002b). During its 1994 Earth flyby, Geographos was going to be visited by the Clementine spacecraft as the secondary mission target after the primary lunar mission was complete (Vorder Bruegge \& Shoemaker 1993). Unfortunately, the spacecraft malfunctioned before leaving the Moon and never reached Geographos.

In addition to these studies, Geographos was observed in the thermal-infrared at 10.1 $\muup$m by Veeder et al. (1989) who obtained two light-curves at this wavelength, and at 12, 25, and 60 $\muup$m by the IRAS satellite in a single snapshot measurement (Green 1985; Tedesco et al. 2004). These observations complete Geographos's data set, and allow a full thermophysical analysis of its Yarkovsky and YORP effects to be performed using the same methodology as that presented in Rozitis et al. (2013) for (1862) Apollo. In the following sections, we present results from application of the advanced thermophysical model (ATPM; Rozitis \& Green 2011, 2012, 2013a), which explicitly incorporates 1D heat conduction, shadowing, multiple scattering of sunlight, global self-heating, and rough surface thermal-infrared beaming, to this complete data set in order to constrain Geographos's thermophysical properties and bulk density.

\section{Thermophysical modelling}

To determine Geographos's thermophysical properties we combine the ATPM with the radar-derived\footnote{\url{http://echo.jpl.nasa.gov/asteroids/shapes/shapes.html}} and light-curve-derived\footnote{\url{http://astro.troja.mff.cuni.cz/projects/asteroids3D/web.php}} shape models and spin states (Hudson \& Ostro 1999; Ďurech et al. 2008b), and compare the model outputs for various thermophysical properties with the thermal-infrared observations obtained from three occasions in 1983 (Green 1985; Veeder et al. 1989; Tedesco et al. 2004) via chi-squared fitting. The methodology used here is exactly the same as that presented in Rozitis et al. (2013). To determine the asteroid thermal emission, the ATPM is used to compute the surface temperature variation for each shape model facet during a rotation by solving the 1D heat conduction equation with a surface boundary condition that includes direct and multiple scattered solar radiation, shadowing, and re-absorbed thermal radiation from interfacing facets. The model explicitly includes rough surface thermal-infrared beaming (i.e. re-radiation of absorbed sunlight back towards the Sun at thermal-infrared wavelengths as a result of surface roughness) from each shape facet by including roughness facets that are arranged in the form of hemispherical craters. The degree of roughness and thermal-infrared beaming is characterised by the fraction of surface,  $f_{\mathrm R}$, covered by the hemispherical craters. A Planck function is applied to the derived temperatures and summed across visible shape and roughness facets to give the emitted thermal flux as a function of wavelength, rotation phase, and various thermophysical properties.

For Geographos, the free parameters to be constrained by fits to the thermal-infrared observations are the effective diameter (i.e. the diameter of an equivalent volume sphere), $D$, geometric albedo, $p_{\mathrm v}$, thermal inertia, $\Gamma$, and surface roughness,  $f_{\mathrm R}$. The effective diameter and geometric albedo are related to the absolute visual magnitude, $H_{\mathrm v}$, by
\begin{equation}
D=\frac{10^{-H_{\mathrm v}/5}1329}{\sqrt{p_{\mathrm v}}}\text{ km}
\end{equation}
and can be considered as a single free parameter (Fowler \& Chillemi 1992). It is also possible to fix the diameter and albedo at the radar-derived values (Hudson \& Ostro 1999), and both possibilities of having a free and fixed diameter will be explored in the model fitting. The model thermal flux predictions, $F_{\text{MOD}}(\lambda_{n}, \varphi_{n},\Gamma,D,f_{\mathrm R})$, were compared with the observations, $F_{\text{OBS}}(\lambda_{n},\varphi_{n})$, and observational errors, $\sigma_{\text{OBS}}(\lambda_{n},\varphi_{n})$, by varying the effective diameter, thermal inertia, and roughness fraction to give the minimum chi-squared fit
\begin{equation}
\chi^{2}=\sum_{n=1}^{N}\Bigg [\frac{FCF(D,f_{\mathrm R})F_{\text{MOD}}(\lambda_{n},\varphi_{n},\Gamma,D,f_{\mathrm R})-F_{\text{OBS}}(\lambda_{n},\varphi_{n})}{\sigma_{\text{OBS}}(\lambda_{n},\varphi_{n})}\Bigg]^{2}
\end{equation}
for a set of $N$ observations with wavelength $\lambda_{n}$ and rotation phase $\varphi_{n}$. The flux correction factor, $FCF(D,f_{\mathrm R})$, is given by
\begin{equation}
FCF(D,f_{\mathrm R})=\frac{1-A_{\mathrm B}(D,f_{\mathrm R})}{1-A_{\text{B\_MOD}}},
\end{equation}
where $A_{\mathrm B}(D,f_{\mathrm R})$ is the required Bond albedo for an asteroid with effective diameter $D$ and roughness fraction $f_{\mathrm R}$, and $A_{\text{B\_MOD}}$ is the model Bond albedo used in the ATPM. This saves a lot of computational effort by not running the model for every value of Bond albedo required, and the flux correction factor is typically within 10\% of unity for an assumed value of $A_{\text{B\_MOD}}=0.06$.

The rotation phase of Geographos at the time of the thermal-infrared observations was calculated using the initial epoch, rotation period, and YORP rotational acceleration determined by Ďurech et al. (2008b) from light-curve inversion. For the radar shape model, an additional rotation offset was required to account for the different co-ordinate systems used in the radar and light-curve shape models. This offset was found by minimising the relative-chi-squared fit (see Eq. (7) of Kaasalainen \& Torppa 2001) of the synthetic radar shape model light-curves to the optical light-curve observations. For comparison purposes, the radar shape model gave a minimised relative-chi-square value of 26.6 whilst the light-curve shape model gave 11.9, which indicates that the light-curve shape model gave (perhaps unsurprisingly) a better fit to the optical light-curve observations.

Separate thermophysical models were run for thermal inertia values ranging from 0 to 3000 J m$^{-2}$ K$^{-1}$ s$^{-1/2}$ in equally spaced steps of 20 J m$^{-2}$ K$^{-1}$ s$^{-1/2}$. Similarly, the effective diameter and roughness fraction were stepped through their plausible ranges, which formed a 3D grid of model test parameters (or test clones) with the thermal inertia steps. A parameter region bounded by a constant $\Delta\chi^{2}$ value at the 3-$\sigma$ confidence value was chosen to define the range of possible parameters, and the three sets of thermal-infrared observations were fitted simultaneously in the ATPM chi-square fitting. Table 1 summarises the two sets of thermal-infrared observations, and Table 2 summarises the fixed model parameters, used to determine Geographos's thermophysical properties.

In the ATPM fitting, the maximum equatorial diameter was initially kept fixed at the Doppler-radar-derived value of 5.04 $\pm$ 0.07 km for both the radar and light-curve shape models, as this was the most reliable diameter measurement available (Ostro et al. 1996). The corresponding effective diameters were then 2.56 $\pm$ 0.03 and 2.46 $\pm$ 0.03 km for the radar and light-curve shape models, respectively. The logic was by using a fixed diameter in the ATPM fitting it would reduce the number of free parameters to 2 (instead of 3), and lead to tighter constraints on the derived thermal inertia and roughness fraction values. However, whilst a good fit (i.e. a reduced chi-squared value of 0.50) was obtained for the light-curve shape model, a bad fit (i.e. a reduced chi-squared value of 2.42) was obtained for the radar shape model using this fixed diameter. A thermal inertia of 340 $_{-100}^{+140}$ J m$^{-2}$ K$^{-1}$ s$^{-1/2}$ was derived for the light-curve shape model whereas a very high thermal inertia of 1320 $_{-440}^{+600}$ J m$^{-2}$ K$^{-1}$ s$^{-1/2}$ was suggested for the radar shape model  (see Table 3). Indeed, the radar shape model value is much greater than the largest thermal inertia derived for any asteroid, i.e. near-Earth asteroid (25143) Itokawa has the highest measured thermal inertia of 750 $_{-300}^{+50}$ J m$^{-2}$ K$^{-1}$ s$^{-1/2}$ (Müller et al. 2005). A much better fit (i.e. a reduced chi-squared value of 0.30) could be obtained for the radar shape model by leaving the diameter as a free parameter in the ATPM fitting. This required a smaller effective diameter of 2.21 $\pm$ 0.13 km and a more consistent thermal inertia of 320 $_{-160}^{+220}$ J m$^{-2}$ K$^{-1}$ s$^{-1/2}$ (see Table 3), which suggested that the radar observations had overestimated the size of Geographos. However, this produced a maximum equatorial diameter of 4.35 $\pm$ 0.26 km, which was inconsistent with that of 5.04 $\pm$ 0.07 km accurately measured by Doppler-radar observations.

Intriguingly, the light-curve shape model with a free diameter still produced similar results to when its diameter was kept fixed in the ATPM fitting  (see Table 3). It also produced a maximum equatorial diameter that agreed with the Doppler-radar observations. This led to suspicions that there was something wrong with the radar shape model. Indeed, by comparing the axial ratios of the two shape models, i.e. 2.51:1.00:1.07 for the radar shape model versus 3.08:1.22:1.00 for the light-curve shape model, it is seen that the light-curve shape model is significantly more oblate than the radar shape model (see Fig. 1). This meant that the radar shape model had more cross-sectional area projected towards the observer, which leads to enhanced model flux for the same effective diameter and thermophysical properties. Therefore, the ATPM fitting tried to counter this by using a larger thermal inertia value to reduce the model flux. It seemed that the radar observations had overestimated the z-axis of Geographos, which is very plausible considering that the radar shape model contains a north-south ambiguity (Hudson \& Ostro 1999).

To check that this was the case, the radar shape model was flattened, i.e. the flattened-radar shape model (see Fig. 1 and Table 2), so that it had the same axial ratios as the light-curve shape model. The radar shape model was flattened by multiplying its z-axis by a factor of 0.767, which reduced the effective diameter (i.e. 2.34 $\pm$ 0.03 km) of Geographos but maintained the same equatorial diameter. Performing the ATPM fitting with the flattened-radar shape model using a fixed diameter produced a much better fit (i.e. a reduced chi-squared value of 0.20) and results that were consistent with the light-curve shape model results, i.e. a thermal inertia of 260 $_{-80}^{+100}$ J m$^{-2}$ K$^{-1}$ s$^{-1/2}$ (see Table 3). Consistent results were also obtained when its diameter was left free in the ATPM fitting too. Furthermore, the flattened-radar shape model also produced a better fit to the optical light-curve observations than the original radar shape model, i.e. a light-curve relative-chi-squared value of 23.3 was obtained versus 26.6 originally. Therefore, simply reducing the z-axis of the radar shape model reconciled the differences seen between the radar and thermal-infrared observations.

Similar levels of surface roughness were derived for the three different shape models. These suggest a roughness greater than 50\%, although the flattened-radar shape model that gave the best fit to the thermal-infrared observations, had the smoothest surface.

Figure 2 gives example ATPM fits to the thermal-infrared observations using the three different shape models with a fixed diameter, and Table 3 summarises the ATPM derived thermophysical properties at the 3-$\sigma$ confidence level. Figure 3 shows the distribution of possible thermal inertia values, and the co-variance of the average roughness fraction with thermal inertia, derived using a fixed diameter. As in Rozitis et al. (2013), the thermal inertia distribution is obtained by counting each allowed test clone with a specific thermal inertia value and dividing by the total number of allowed test clones ($\sim$10$^{4}$ clones). The co-variance of the roughness fraction with thermal inertia is obtained by averaging the values of the allowed test clones in each thermal inertia bin (n.b. this format is also used in the Yarkovsky and YORP effect predictions presented later in this work).

\section{Yarkovsky and YORP modelling}

The Yarkovsky and YORP effects acting on Geographos can be determined by computing the total recoil forces and torques from reflected and thermally emitted photons from the asteroid surface (see Rozitis \& Green 2012, 2013a, for methodology). The inclusion of rough surface thermal-infrared beaming effects in the predictions, on average, enhances the Yarkovsky orbital drift whilst it dampens the YORP rotational acceleration by orders of several tens of per cent (Rozitis \& Green 2012). Including global self-heating effects does not significantly affect the Yarkovsky orbital drift predictions (less than a few per cent difference) but can significantly affect the YORP rotational acceleration predictions in some cases (Rozitis \& Green 2013a). Utilising the three shape models of Geographos, Yarkovsky and YORP effect predictions were made for the range of possible thermophysical properties (or allowed test clones) determined by the thermal-infrared flux fitting and compared against published measurements.

Like (1862) Apollo in Rozitis et al. (2013), Geographos has three measurements of Yarkovsky semimajor axis drift: -17.7 $\pm$ 5.9 m yr$^{-1}$ by Chesley et al. (2008), -37.4 $\pm$ 9.0 m yr$^{-1}$ by Nugent et al. (2012), and -27.2 $\pm$ 9.0 m yr$^{-1}$ by Farnocchia et al. (2013). We combined the three measurements to produce an average drift of -27.4 m yr$^{-1}$ with a standard error of 5.7 m yr$^{-1}$, which we used to produce a normal distribution of possible drifts that has a mean and standard deviation equal to these values. For model comparison, 200 values of possible Yarkovsky drift were randomly selected from this distribution that ranged from -14.0 to -43.0 m yr$^{-1}$, which ensured that the three measured values were encompassed (see Fig. 4). The overall predicted Yarkovsky drift,  $\mathrm d a/\mathrm d t(\Gamma,D,f_{\mathrm R},\rho)$, for a bulk density of $\rho$ was determined from the predicted smooth surface drift, $\mathrm d a/\mathrm d t(\Gamma)_{\text{smooth}}$, the rough surface drift, $\mathrm d a/\mathrm d t(\Gamma)_{\text{rough}}$, and the seasonal drift, $\mathrm d a/\mathrm d t(\Gamma)_{\text{seasonal}}$, using
\begin{equation}
\begin{split}
\frac{\mathrm d a}{\mathrm d t}(\Gamma,D,f_{\mathrm R},\rho)=&\Bigg(\frac{D_{0}}{D}\Bigg)\Bigg(\frac{\rho_{0}}{\rho}\Bigg)FCF(f_{\mathrm R})\Bigg[(1-f_{\mathrm R})\frac{\mathrm d a}{\mathrm d t}(\Gamma)_{\text{smooth}}\\
&+f_{\mathrm R}\frac{\mathrm d a}{\mathrm d t}(\Gamma)_{\text{rough}}+\frac{\mathrm d a}{\mathrm d t}(\Gamma)_{\text{seasonal}}\Bigg],
\end{split}
\end{equation}
where each component had been evaluated separately using the ATPM at an initial diameter $D_{0}$ and bulk density $\rho_{0}$. It was model convenient to treat the three components separately this way (see Rozitis \& Green 2012, 2013a for more details), as each component arose from surface temperature distributions occurring at different spatial scales (i.e. shape facet scale for smooth component, roughness facet scale for rough component, and seasonal thermal wave scale for seasonal component). The bulk density for a set of properties, $\rho(\Gamma,D,f_{\mathrm R})$, could then be determined from the measured Yarkovsky drift, $\mathrm d a/\mathrm d t_{\text{measured}}$, using
\begin{equation}
\rho(\Gamma,D,f_{\mathrm R})=\rho_{0}\Bigg(\frac{\mathrm d a}{\mathrm d t}(\Gamma,D,f_{\mathrm R},\rho_{0})\Bigg/\frac{\mathrm d a}{\mathrm d t}_{\text{measured}}\Bigg)\text{ .}
\end{equation}

Yarkovsky/YORP modelling and bulk density determination were peformed for both a free and fixed diameter, however, the fixed diameter modelling was taken as the nominal result. Figure 5a shows the average Yarkovsky drift as a function of thermal inertia with fixed bulk density. Figure 5b shows the average bulk density required to match the observed orbital drift as a function of thermal inertia. Lastly, Fig. 5c shows the distribution of possible bulk densities derived separately for the three different shape models using a fixed diameter. The derived bulk densities have median values and 1-$\sigma$ spreads of 1450 $_{-350}^{+450}$, 2150 $_{-400}^{+600}$, and 2100 $_{-450}^{+550}$ kg m$^{-3}$ for the radar, flattened-radar, and light-curve shape models, respectively. Whilst the flattened-radar and light-curve shape models produced very consistent results, the radar shape model produced a much lower bulk density because of its much higher thermal inertia value when derived using a fixed diameter. However, the radar shape model produced a more consistent bulk density of 2400 $_{-500}^{+650}$ kg m$^{-3}$ when using a free diameter. A free diameter also produced consistent results for the flattened-radar and light-curve shape models but with slightly larger uncertainties when compared to the fixed diameter results.

Ďurech et al. (2008b) has measured Geographos's YORP rotational acceleration to be (1.5 $\pm$ 0.2) $\times10^{-3}$ rad yr$^{-2}$, which we use for model comparison. As in Rozitis et al. (2013), the overall YORP rotational acceleration, $\mathrm d \omega/\mathrm d t(D,f_{\mathrm R},\rho)$, can be predicted using
\begin{equation}
\frac{\mathrm d \omega}{\mathrm d t}(D,f_{\mathrm R},\rho)=\Bigg(\frac{D_{0}}{D}\Bigg)^{2}\Bigg(\frac{\rho_{0}}{\rho}\Bigg)\Bigg[(1-f_{\mathrm R})\frac{\mathrm d \omega}{\mathrm d t}_{\text{smooth}}+f_{\mathrm R}\frac{\mathrm d \omega}{\mathrm d t}_{\text{rough}}\Bigg],
\end{equation}
where  $\mathrm d \omega/\mathrm d t_{\text{smooth}}$ and  $\mathrm d \omega/\mathrm d t_{\text{rough}}$ are the YORP rotational acceleration values for a smooth and rough surface, respectively, which are independent of thermal inertia. Similarly, the rate of YORP obliquity shift,  $\mathrm d \xi/\mathrm d t(\Gamma,D,f_{\mathrm R},\rho)$, can be predicted using
\begin{equation}
\frac{\mathrm d \xi}{\mathrm d t}(\Gamma,D,f_{\mathrm R},\rho)=\Bigg(\frac{D_{0}}{D}\Bigg)^{2}\Bigg(\frac{\rho_{0}}{\rho}\Bigg)\Bigg[(1-f_{\mathrm R})\frac{\mathrm d \xi}{\mathrm d t}(\Gamma)_{\text{smooth}}+f_{\mathrm R}\frac{\mathrm d \xi}{\mathrm d t}(\Gamma)_{\text{rough}}\Bigg],
\end{equation}
where $\mathrm d \xi/\mathrm d t(\Gamma)_{\text{smooth}}$ and $\mathrm d \xi/\mathrm d t(\Gamma)_{\text{rough}}$ are the rates of YORP obliquity shift for a smooth and rough surface, respectively, which are dependent on thermal inertia in this case. 

The bulk density values used here were those determined by the model-to-measurement comparisons of the Yarkovsky drift from Eq. (4). Whilst it was conceivable to try and fit the bulk density using model-to-measurement comparisons of both the Yarkovsky drift and YORP rotational acceleration simultaneously, it was not done so here because the error in the YORP effect prediction could be very large (see section 4.1). Figure 6a shows the average YORP rotational acceleration as a function of thermal inertia, and Fig. 6b shows the distribution of possible YORP rotational accelerations derived separately for the three different shape models. The derived YORP rotational accelerations using a fixed diameter have median values and 1-$\sigma$ spreads of (-5.5 $_{-1.6}^{+1.3}$), (-4.2 $_{-1.3}^{+1.0}$), and (1.9 $_{-0.4}^{+0.5}$) $\times10^{-3}$ rad yr$^{-2}$ for the radar, flattened-radar, and light-curve shape models, respectively. Only the light-curve shape model produced a prediction range that agreed with the measured value of (1.5 $\pm$ 0.2) $\times10^{-3}$ rad yr$^{-2}$, as the radar and flattened-radar shape models produced predictions with opposite sign to that observed. Figure 6c shows the average rate of YORP obliquity shift, and Fig. 6d shows the distribution of possible values. The derived rates of YORP obliquity shift have median and 1-$\sigma$ spreads of 1.0 $\pm$ 0.4, 4.2 $_{-1.2}^{+1.5}$, and 1.9 $_{-0.4}^{+0.6}$ degrees per $10^{5}$ yr for the radar, flattened-radar, and light-curve shape models, respectively. Again, a free diameter produced consistent results but with slightly larger uncertainties. Table 4 summarises Geographos's density, mass, and spin change properties derived for the three different shape models investigated.

\section{Discussion}
\subsection{Modelling critiques}

The accuracy of the Yarkovsky and YORP effect analysis described in section 3 depends strongly on the accuracy of the measured Yarkovsky semimajor axis drift, and for Geographos there are three different measurements giving three slightly different values. It is not clear where these differences arise but it will depend on the parameter fitting, the data weighting treament, and the data sets used in their respective astrometric studies. Fortunately, these measurement differences are not significant enough for them to be considered inconsistent with one another. For example, the largest difference arises between the Chesley et al. (2008) measurement and the Nugent et al. (2012) measurement, which is 1.8-$\sigma$ using their quoted uncertainties. The Farnocchia et al. (2013) measurement is consistent with the two other measurements at the 1-$\sigma$ level. For us to assume that they are inconsistent would require their differences to be greater than the 3-$\sigma$ level, which is clearly not the case here. In the nominal Yarkovsky and YORP effect analysis described above, Monte Carlo sampling from a normal distribution derived from all three measurements was utilised in an attempt to combine the three measurements in a statistically meaningful way (see Fig. 4). However, if one of the measurements could be rejected for a valid reason then the results of the analysis would change. The only reason to consider this is for removal of the Chesley et al. (2008) measurement, as it could be considered to be superseded by the updated analysis described in Nugent et al. (2012) and Farnocchia et al. (2013). A higher rate of Yarkovsky drift would then be preferred, which would lower the derived bulk density and increase the predicted value of the YORP rotational acceleration and obliquity shift (because $\mathrm d \omega/\mathrm d t\propto1 / \rho\propto \mathrm d a/\mathrm d t$).

In the YORP effect modelling, it was the light-curve shape model that provided the best fit to the observations rather than the radar and flattened-radar shape models. In particular, the radar and flattened-radar shape models could not predict the correct sign of the YORP rotational acceleration. The opposite sign YORP rotational acceleration was also found by Ďurech et al. (2008b), and Rozitis \& Green (2012) speculated that this is because the Hudson \& Ostro (1999) shape model is not unique as the data set it was derived from contains north-south ambiguities due to the near-equatorial view of the asteroid during the radar observations. The light-curve observations were taken from multiple and different geometries, and the resulting shape model produced by light-curve inversion contains no significant degeneracy. As demonstrated, flattening the radar shape model by reducing its z-axis did not reconcile the opposite sign prediction because it equally affected YORP driving shape features located on opposite sides of the asteroid. Reconciliation would require these shape features to be modified unevenly.   

Like (1862) Apollo in Rozitis et al. (2013), it is surprising that the theoretical YORP rotational acceleration predicted by the light-curve shape model agrees with the observed value quite well. Especially when previous studies have shown that the YORP effect can be highly sensitive to unresolved shape features and surface roughness (Statler 2009; Rozitis \& Green 2012), the shape model resolution (Breiter et al. 2009), and internal bulk density distribution (Scheeres \& Gaskell 2008; Lowry et al. 2014). However, Geographos has a relatively high YORP-coefficient of ~0.01 (see Rossi et al. 2009 or Rozitis \& Green 2013b for a definition), and Rozitis \& Green (2013a) showed that asteroids with high values are less sensitive to the inclusion of concavities in their global shape model. Furthermore, the Geographos prediction is relatively insensitive to small-scale shape features, as when the roughness is allowed to vary in an extreme way across the surface it only introduced an uncertainty of $\sim$30\%. Similar findings were made by Ďurech et al. (2008b) and Kaasalainen \& Nortunen (2013). In Ďurech et al. (2008b), they added small-scale topography from the spacecraft-derived and high-resolution shape model of (25143) Itokawa (Gaskell 2008) to the light-curve shape model of Geographos and found no differences larger than $\sim$5\% between their YORP rotational acceleration predictions. In Kaasalainen \& Nortunen (2013), they find that the YORP rotational acceleration prediction produced by the light-curve shape model of Geographos is stable and semi-stable against local and global shape perturbations respectively.

\subsection{Reconciling radar and thermal-infrared observations}

The poor fit of the radar shape model with its nominal diameter highlights the importance of having a good shape model for thermophysical modelling of irregular shaped asteroids. Not every shape model, whether it is radar-derived or light-curve-derived, produces results that are consistent with other types of observational data. During shape model inversion there is generally a range of similarly-looking shapes that fit equally well to a set of data of one type, and one nominal solution is usually given in publications. In certain circumstances, such as when there is limited data or the range of viewing geometries is small, acceptable shape models can differ significantly. This mainly manifests itself through a large uncertainty in the shape model z-axis.

For example, radar observations in some cases cannot determine whether an asteroid is a prograde or a retrograde rotator and there are two possible shape models, which is the case for near-Earth asteroids (4486) Mithra (Brozovic et al. 2010) and (29075) 1950 DA (Busch et al. 2007). In other cases, radar-derived shape models sometimes contain a north-south ambiquity because of a near-equatorial view during the time of the radar observations. Depending on the acquired data set, the resulting uncertainty in the shape model z-axis could be very large, e.g. as seen for Geographos (Hudson \& Ostro 1999), or relatively small, e.g. as seen for (101955) Bennu (Nolan et al. 2013). Similar issues can also affect shape models derived from light-curve observations. For example, Ďurech et al. (2008a) noted that the z-axis of the (1862) Apollo shape model was not well constrained from the light-curve photometry, and their nominal shape model appeared to be too flat. Lowry et al. (2012) also demonstrated that even with a good light-curve data set there can be some flexibility in the derived shape model, which resulted in an uncertainty of $\sim$7\% in the axial ratios determined for the nucleus of comet 67P/Churymov-Gerasimenko.

Generally, Doppler-radar observations provide the most accurate measurement of an asteroid's size through determination of its equatorial extent (Ostro et al. 2002). This is because the returned Doppler-broadened signal depends only on the asteroids's equatorial diameter, rotation period, and pole orientation. If the latter two asteroid properties are well known then the equatorial diameter can be accurately measured, and a shape model can be scaled to have this value. However, if the shape model z-axis is overestimated then the effective diameter is also overestimated despite it having the correct equatorial diameter. Likewise, if the shape model z-axis is underestimated then the effective diameter will also be underestimated. Thermal-infrared observations are more sensitive to the effective diameter rather than the equatorial diameter, as it is the cross-sectional area projected towards the observer that is important in this case. If the spatial extent in the shape model z-axis is wrong then there will be a mis-match between the radar-derived and thermal-derived effective diameters. This was demonstrated to be the case here for Geographos, which had an overestimated z-axis for its radar shape model. It was also previously demonstrated for (1862) Apollo, which in that case had an underestimated z-axis in its light-curve-derived shape model (Rozitis et al. 2013). Large uncertainties in shape model z-axes are likely to explain why large diameter differences were obtained for near-Earth asteroids 2002 NY40 (Müller et al. 2004) and (308635) 2005 YU55 (Müller et al. 2013). It could potentially explain why some thermophysical model fits have high residuals despite producing consistent diameters, e.g. as seen for (341843) 2008 EV5 (Alí-Lagoa et al. 2014), because the error in the z-axis also alters the modelled surface temperatures away from the correct ones.

Ideally, the shape inversion and thermophysical modelling should be done simultaneously such that both aspects are optimised, as was done in a rudimentary way for (1862) Apollo in Rozitis et al. (2013). However, the combined modelling required might be too complex for current computational hardware but should be investigated in future work (e.g. Ďurech et al. 2012b).

\subsection{Further implications for Geographos}

The macro-porosity for Geographos can be estimated by assuming a typical bulk density of $\sim$3330 kg m$^{-3}$ for the ordinary chondrites that are associated with S-type asteroids (Carry 2012). The bulk densities of 1450 $_{-350}^{+450}$, 2150 $_{-400}^{+600}$, and 2100 $_{-450}^{+550}$ kg m$^{-3}$ derived using a fixed diameter for the radar, flattened-radar, and light-curve shape models give macro-porosities of 56 $_{-13}^{+11}$, 35 $_{-18}^{+12}$, and 37 $_{-17}^{+13}$ \%, respectively. These values indicate that Geographos most likely has a rubble-pile interior structure (Britt et al. 2002). This conclusion would remain valid if the Chesley et al. (2008) Yarkovsky drift measurement was rejected (see section 4.1), as the bulk density would be lowered and the macro-porosity would be increased (to $\sim$1800 kg m$^{-3}$ and $\sim$46\% respectively) to take into account the increased rate of Yarkovsky drift (relative to our nominal value) of Geographos. The rubble-pile nature is consistent with its irregular and highly elongated shape having formed from tidal distortion during a close planetary encounter (Solem \& Hills 1996; Bottke et al. 1999). The measured bulk density is also lower than 2400 kg m$^{-3}$ assumed in the meteoroid ejection studies (Ryabova 2002a,b), which suggests that it is easier for loose material (as inferred from its thermal inertia value below) to be lofted away from Geographos's surface during close encounters with the Earth than previously thought.

Taking the light-curve shape model with a fixed diameter as the nominal result, Geographos's thermal inertia of 340 $_{-100}^{+140}$ J m$^{-2}$ K$^{-1}$ s$^{-1/2}$ is slightly high for a near-Earth asteroid that is a few km in size. In particular, it is larger than 180 $\pm$ 50, 140 $_{-100}^{+140}$, and 120 $\pm$ 50 J m$^{-2}$ K$^{-1}$ s$^{-1/2}$ determined for the near-Earth asteroids (1580) Betulia (Harris et al. 2005), (1862) Apollo (Rozitis et al. 2013), and (175706) 1996 FG3 (Wolters et al. 2011), respectively. It is also larger than the mean value of 200 $\pm$ 40 J m$^{-2}$ K$^{-1}$ s$^{-1/2}$ determined for km-sized near-Earth asteroids (Delbo et al. 2007). It is most similar to that of 310 $\pm$ 70 and 400 $\pm$ 200 J m$^{-2}$ K$^{-1}$ s$^{-1/2}$ determined for the sub-km near-Earth asteroids (101955) Bennu (Emery et al. 2014) and (162173) 1999 JU3 (Müller et al. 2011), respectively. Its surface is therefore likely to consist of a mixture of fine grains and large rocks/boulders rather than just fine grains. The derived bulk density of 2100 $_{-450}^{+550}$ kg m$^{-3}$ is very similar to that of 1950 $\pm$ 140 determined for the S-type near-Earth asteroid (25143) Itokawa (Abe et al. 2006). This suggests that Geographos could have a very similar interior structure to (25143) Itokawa despite being significantly larger, i.e. 2.46 km versus 0.33 km in effective diameter (Fujiwara et al. 2006).

Finally, the obliquity of Geographos is increasing for all three shape models, which indicates that it is approaching one of the YORP asymptotic states at 180\degr~obliquity (Čapek \& Vokrouhlický 2004). Like (1862) Apollo (Rozitis et al. 2013), the YORP effect will halve Geographos's rotation period and shift its rotation axis to the 180\degr~obliquity asymptotic state in just $\sim$7 Myr, whilst in the same amount of time the Yarkovsky effect will decrease Geographos's semimajor axis by just $\sim$10$^{-3}$ AU. Therefore, the YORP effect will dominate Geographos's long term evolution.

\begin{acknowledgements}
The authors acknowledge the financial support of the UK Science and Technology Facilities Council (STFC), and are grateful to the reviewer Marco Delbo for several suggested improvements to the paper.
\end{acknowledgements}

\onecolumn

\begin{table}
\caption{Summary of the (1620) Geographos thermal-infrared observations obtained in 1983.}
\centering
\begin{tabular}{l c c c c c}
\hline\hline
Observation & Wavelength & Flux ($10^{-14}$ & Heliocentric & Geocentric & Phase angle\\
date (1983 UT) & ($\muup$m) & W m$^{-2}$ $\muup$m$^{-1}$) & distance (AU) & distance (AU) & (\degr)\\
\hline
March 06.36\tablefootmark{a} & 10.1 & 6.25 $\pm$ 0.58 & & & \\
March 06.41\tablefootmark{a} & 10.1 & 23.6 $\pm$ 2.2 & 1.098 & 0.111 & 16.9 \\
March 06.47\tablefootmark{a} & 10.1 & 8.32 $\pm$ 0.77 & & & \\
March 11.31\tablefootmark{a} & 10.1 & 27.1 $\pm$ 2.1 & & & \\
March 11.36\tablefootmark{a} & 10.1 & 7.25 $\pm$ 0.67 & 1.071 & 0.095 & 34.0 \\
March 11.42\tablefootmark{a} & 10.1 & 25.8 $\pm$ 2.4 & & & \\
March 11.47\tablefootmark{a} & 10.1 & 8.25 $\pm$ 0.76 & & & \\
March 24.10\tablefootmark{b} & 12 & 3.20 $\pm$ 0.42 & & & \\
March 24.10\tablefootmark{b} & 25 & 1.29 $\pm$ 0.19 & 1.003 & 0.101 & 83.8 \\
March 24.10\tablefootmark{b} & 60 & 0.125 $\pm$ 0.026 & & & \\
\hline
\end{tabular}
\tablefoot{Obtained from \tablefoottext{a}{Veeder et al. (1989)}, and \tablefoottext{b}{Green (1985) and Tedesco et al. (2004)}. The IRAS fluxes quoted here were obtained from Tedesco et al. (2004) and colour corrected using the corrections of Green (1985) with an assumed temperature of 330 K. These corrections were 1.05, 0.85, and 0.85 for the 12, 25, and 60 $\muup$m fluxes, respectively.}
\end{table}

\begin{table}
\caption{Assumed and previously measured thermophysical modelling parameters for thermal-infrared flux fitting and Yarkovsky and YORP effect modelling.}
\centering
\begin{tabular}{l c c c}
\hline\hline
Shape model & Radar & Flattened-radar & Light-curve \\
\hline
Number of vertices & 2048\tablefootmark{a} & 2048\tablefootmark{a} & 1022\tablefootmark{b} \\
Number of facets & 4092\tablefootmark{a} &  4092\tablefootmark{a} & 2040\tablefootmark{b} \\
Axis ratios ($a$:$b$:$c$) & 2.51:1.00:1.07\tablefootmark{a} & 3.07:1.22:1.00 & 3.08:1.22:1.00\tablefootmark{b} \\
Rotation period\tablefootmark{b} & 5.223336 hrs & 5.223336 hrs & 5.223336 hrs \\
YORP rotational acceleration\tablefootmark{b} & (1.5 $\pm$ 0.2) $\times10^{-3}$ rad yr$^{-2}$ & (1.5 $\pm$ 0.2) $\times10^{-3}$ rad yr$^{-2}$ & (1.5 $\pm$ 0.2) $\times10^{-3}$ rad yr$^{-2}$ \\
Pole orientation & $\lambda\tablefootmark{a}=55\degr, \beta\tablefootmark{a}=-46\degr$ & $\lambda\tablefootmark{a}=55\degr, \beta\tablefootmark{a}=-46\degr$ & $\lambda\tablefootmark{b}=58\degr, \beta\tablefootmark{b}=-49\degr$ \\
Obliquity & 149\degr & 149\degr & 152\degr \\
Semimajor axis\tablefootmark{c} & 1.246 AU & 1.246 AU & 1.246 AU \\
Eccentricity\tablefootmark{c} & 0.336 & 0.336 & 0.336 \\
Yarkovsky semimajor axis drift\tablefootmark{d} & -27.4 $\pm$ 5.7 m yr$^{-1}$ & -27.4 $\pm$ 5.7 m yr$^{-1}$ & -27.4 $\pm$ 5.7 m yr$^{-1}$ \\
Absolute magnitude\tablefootmark{e} & 15.6 $\pm$ 0.1 & 15.6 $\pm$ 0.1 & 15.6 $\pm$ 0.1 \\
Phase parameter\tablefootmark{e} & 0.15 & 0.15 & 0.15 \\
Emissivity & 0.9 & 0.9 & 0.9 \\
\hline
\end{tabular}
\tablefoot{Obtained from \tablefoottext{a}{Hudson \& Ostro (1999)}, \tablefoottext{b}{Ďurech et al. (2008b)}, \tablefoottext{c}{the JPL Small-Body Database Browser}, \tablefoottext{d}{Chesley et al. (2008), Nugent et al. (2012), and Farnocchia et al. (2013)}, and \tablefoottext{e}{Magnusson et al. (1996)}.}
\end{table}

\begin{table}
\caption{ATPM derived thermophysical properties of (1620) Geographos using the thermal-infrared observations obtained in 1983 at the 3-$\sigma$ confidence level.}
\centering
\begin{tabular}{l c c c c c c}
\hline\hline
Shape model & \multicolumn{2}{c}{Radar} & \multicolumn{2}{c}{Flattened-radar} & \multicolumn{2}{c}{Light-curve} \\
Diameter fitting mode & Free & Fixed & Free & Fixed & Free & Fixed \\
\hline
Thermal-infrared reduced-$\chi^{2}$ & 0.30 & 2.42 & 0.23 & 0.20 & 0.53 & 0.50 \\
Light-curve relative-$\chi^{2}$ & 26.6 & 26.6 & 23.3 & 23.3 & 11.9 & 11.9 \\
Effective diameter\tablefootmark{a} (km) & 2.21 $\pm$ 0.13 & 2.56 $\pm$ 0.03 & 2.42 $\pm$ 0.13 & 2.34 $\pm$ 0.03 & 2.36 $\pm$ 0.12 & 2.46 $\pm$ 0.03 \\
Maximum equatorial diameter\tablefootmark{a} (km) & 4.35 $\pm$ 0.26 & 5.04 $\pm$ 0.07 & 5.21 $\pm$ 0.28 & 5.04 $\pm$ 0.07 & 4.83 $\pm$ 0.25 & 5.04 $\pm$ 0.07 \\
Geometric albedo\tablefootmark{a} & 0.21 $\pm$ 0.04 & 0.155 $\pm$ 0.015 & 0.17 $\pm$ 0.03 & 0.186 $\pm$ 0.019 & 0.18 $\pm$ 0.03 & 0.168 $\pm$ 0.017 \\
Thermal inertia\tablefootmark{b} (J m$^{-2}$ K$^{-1}$ s$^{-1/2}$) & 320 $_{-160}^{+220}$  & 1320 $_{-440}^{+600}$ & 340 $_{-160}^{+220}$ & 260 $_{-80}^{+100}$ & 260 $_{-120}^{+180}$ & 340 $_{-100}^{+140}$ \\
Roughness fraction (\%) & 60 $\pm$ 26\tablefootmark{a} & $\geq$ 60\tablefootmark{c} & 64 $\pm$ 24\tablefootmark{a} & 58 $\pm$ 24\tablefootmark{a} & $\geq$ 45\tablefootmark{c} & $\geq$ 50\tablefootmark{c} \\
\hline
\end{tabular}
\tablefoot{\tablefoottext{a}{Mean and standard error.}\tablefoottext{b}{Median and 1-$\sigma$ spread.}\tablefoottext{c}{3-$\sigma$ range.}}
\end{table}

\begin{table}
\caption{Mass and spin change properties of (1620) Geographos derived by ATPM at the 3-$\sigma$ confidence level.}
\label{Table 4}
\centering
\begin{tabular}{l c c c c c c}
\hline\hline
Shape model & \multicolumn{2}{c}{Radar} & \multicolumn{2}{c}{Flattened-radar} & \multicolumn{2}{c}{Light-curve} \\
Diameter fitting mode & Free & Fixed & Free & Fixed & Free & Fixed \\
\hline
Bulk density\tablefootmark{a} (kg m$^{-3}$) & 2400 $_{-500}^{+650}$ & 1450 $_{-350}^{+450}$ & 2050 $_{-450}^{+550}$ & 2150 $_{-400}^{+600}$ & 2150 $_{-450}^{+600}$ & 2100 $_{-450}^{+550}$ \\
Macro-porosity\tablefootmark{a} (\%) & 28 $_{-20}^{+15}$ & 56 $_{-13}^{+11}$ & 38 $_{-16}^{+14}$ & 35 $_{-18}^{+12}$ & 35 $_{-18}^{+14}$ & 37 $_{-17}^{+13}$ \\
Mass\tablefootmark{a} ($10^{13}$ kg) & 1.38 $_{-0.21}^{+0.38}$ & 1.25 $_{-0.30}^{+0.39}$ & 1.54 $_{-0.34}^{+0.43}$ & 1.49 $_{-0.31}^{+0.38}$ & 1.48 $_{-0.32}^{+0.42}$ & 1.62 $_{-0.33}^{+0.39}$ \\
Moment of inertia\tablefootmark{a} ($10^{19}$ kg m$^{2}$) & 1.42 $_{-0.40}^{+0.50}$ & 1.67 $_{-0.38}^{+0.53}$ & 2.27 $_{-0.64}^{+0.78}$ & 2.03 $_{-0.43}^{+0.57}$ & 1.91 $_{-0.52}^{+0.66}$ & 2.21 $_{-0.46}^{+0.58}$ \\
YORP rotational acceleration\tablefootmark{a} ($10^{-3}$ rad yr$^{-2}$) &-4.8 $_{-1.6}^{+1.1}$ & -5.5 $_{-1.6}^{+1.3}$ & -4.1 $_{-1.3}^{+1.0}$ & -4.2 $_{-1.3}^{+1.0}$ & 2.0 $_{-0.5}^{+0.6}$ & 1.9 $_{-0.4}^{+0.5}$ \\
YORP obliquity shift\tablefootmark{a} (\degr~/ $10^{5}$ yr) & 2.6 $_{-1.0}^{+1.6}$ & 1.0 $\pm$ 0.4 & 3.6 $_{-1.3}^{+1.8}$ & 4.2 $_{-1.2}^{+1.5}$ & 2.5 $_{-0.6}^{+0.9}$ & 2.3 $\pm$ 0.6 \\
\hline
\end{tabular}
\tablefoot{\tablefoottext{a}{Median and 1-$\sigma$ spread.}}
\end{table}

\begin{figure}
\centering
\includegraphics[width=\hsize]{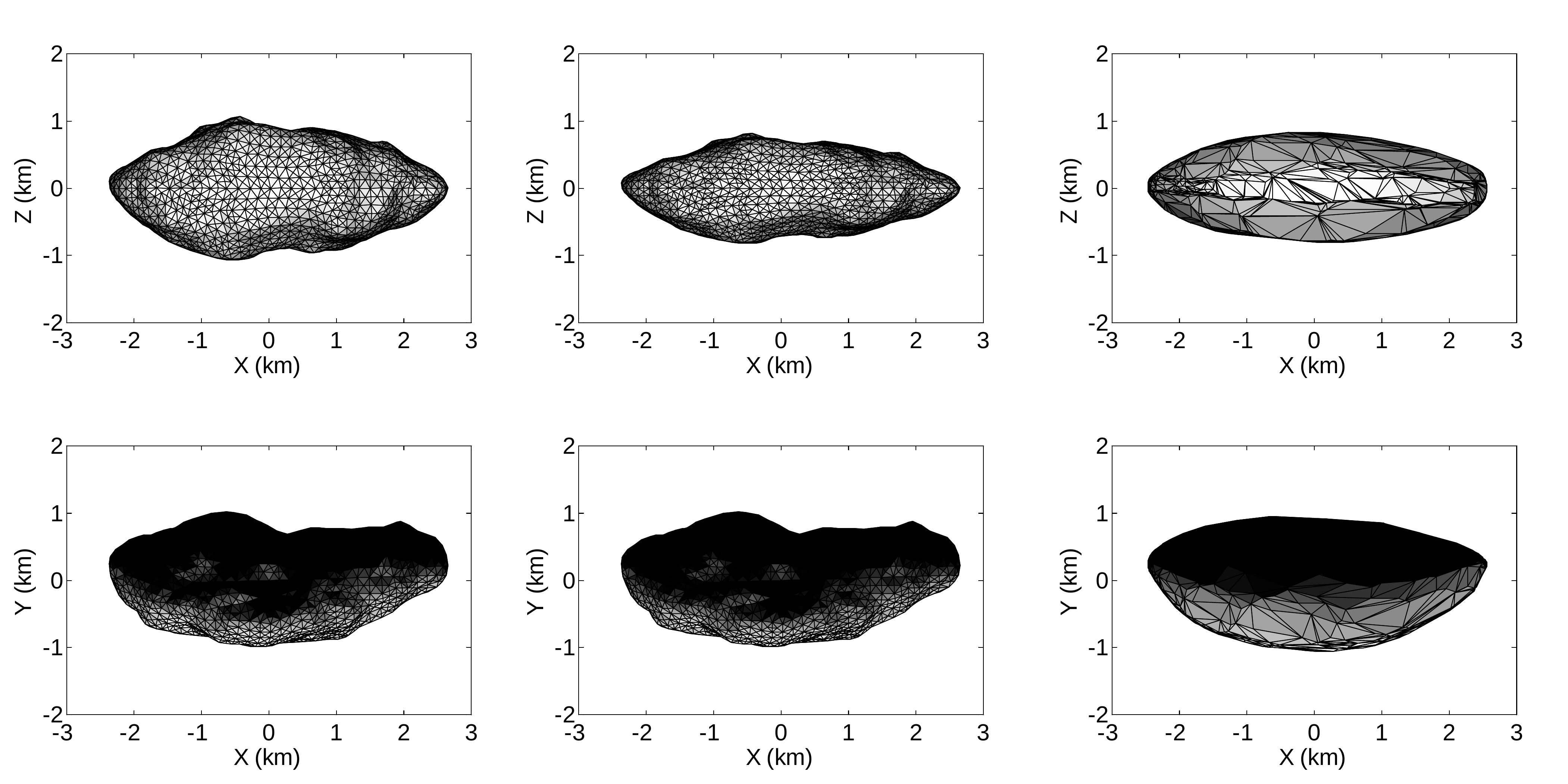}
\caption{The radar (left column; Hudson \& Ostro 1999), flattened-radar (middle column), and light-curve (right column; Ďurech et al. 2008b) shape models of (1620) Geographos. The shape models have been scaled so that they have the same maximum equatorial diameter of 5.04 km, and they have been illuminated by the Sun along their y-axes.}
\end{figure}

\begin{figure}
\centering
\includegraphics[width=\hsize]{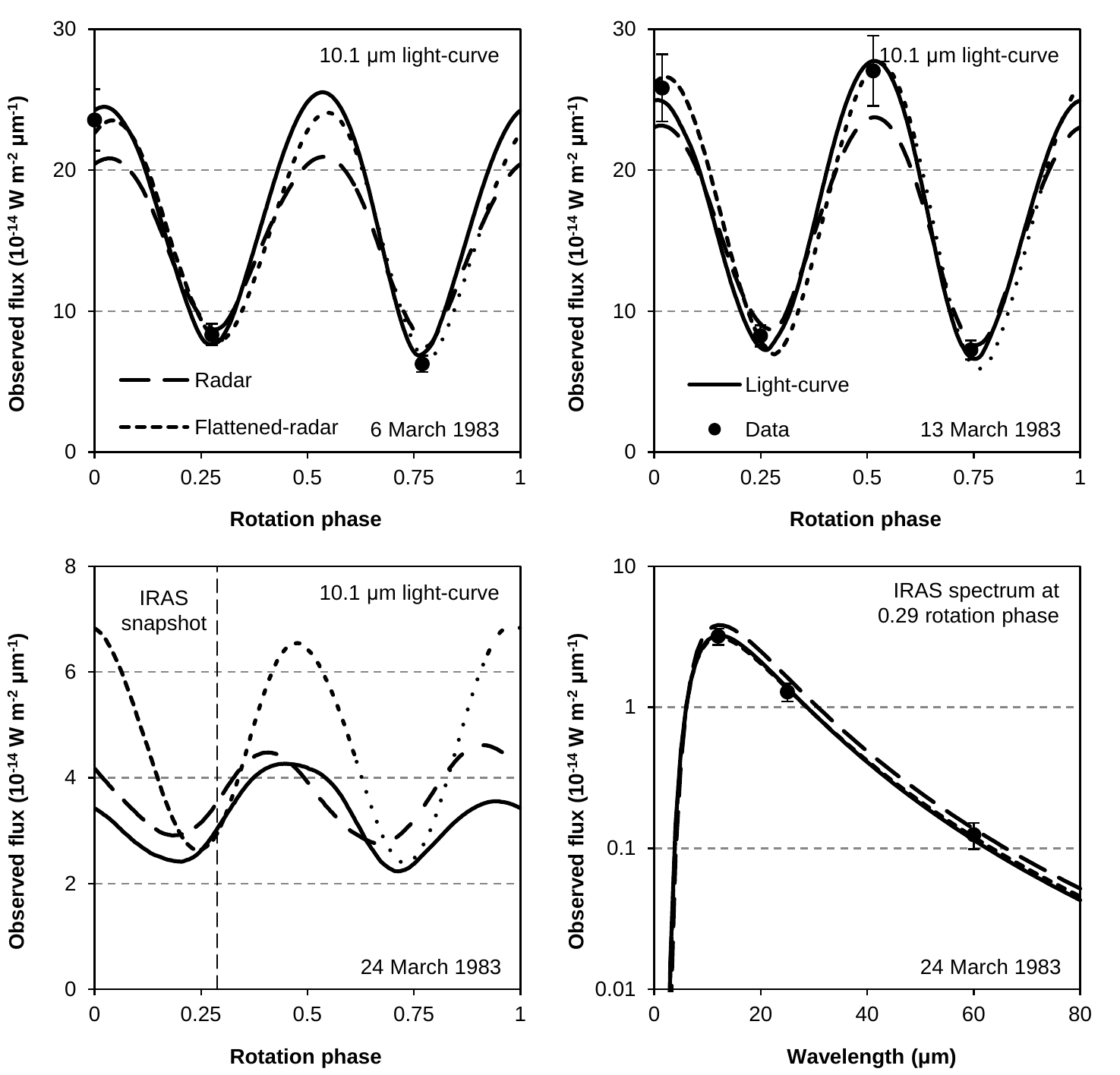}
\caption{Example ATPM fits to the (1620) Geographos thermal-infared observations (data points; Green 1985; Veeder et al. 1989; Tedesco et al. 2004) using the radar (dashed lines), flattened-radar (dotted lines), and light-curve (solid lines) shape models. The thermal-infrared spectrum of the flattened-radar shape model in the lower right panel is almost identical to that of the light-curve shape model.}
\end{figure}

\begin{figure}
\centering
\includegraphics[width=\hsize]{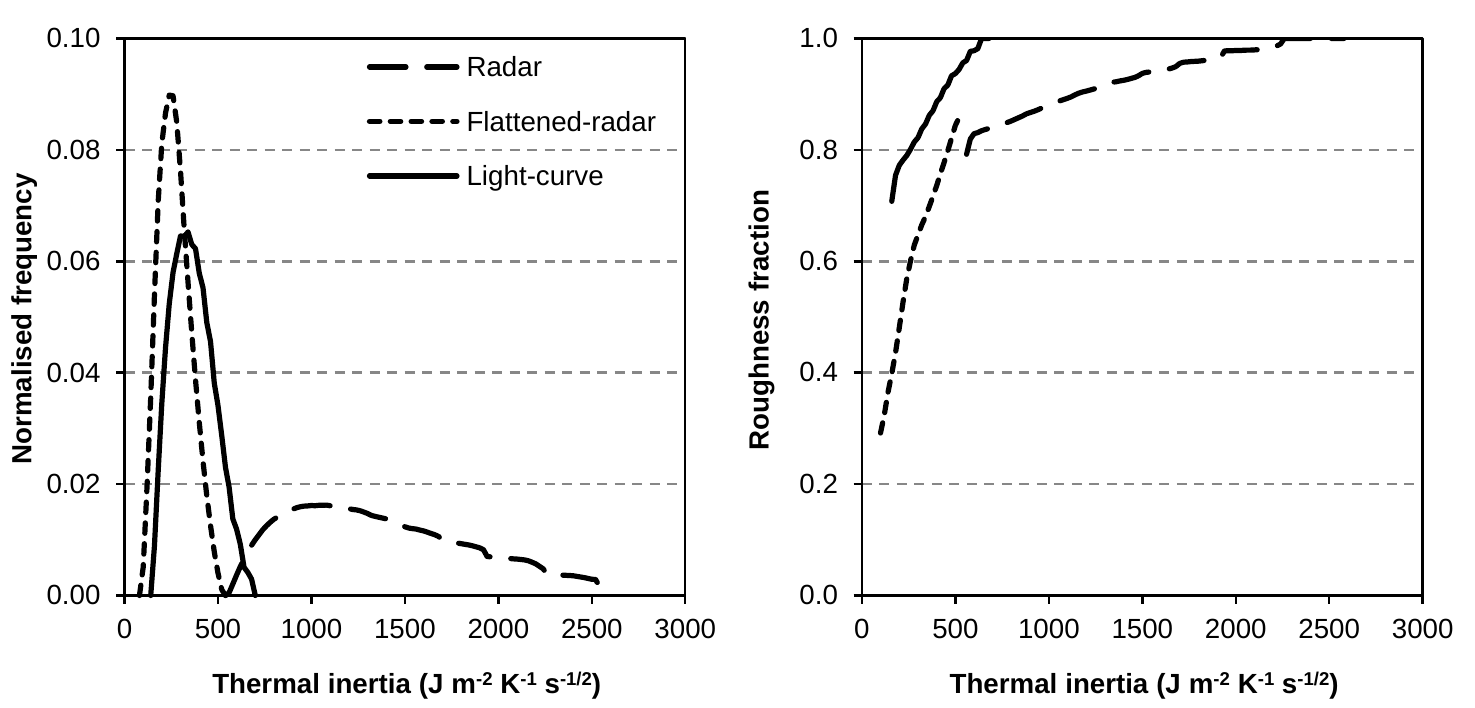}
\caption{Summary of ATPM chi-square fitting results to the thermal-infrared observations of (1620) Geographos obtained in 1983 using a fixed maximum equatorial diameter of 5.04 $\pm$ 0.07 km. The possible thermal inertia distribution (left) and the co-variance with thermal inertia of the average roughness fraction (right) derived at the 3-$\sigma$ confidence level for the three different shape models (legend).}
\end{figure}

\begin{figure}
\centering
\includegraphics[width=\hsize/2]{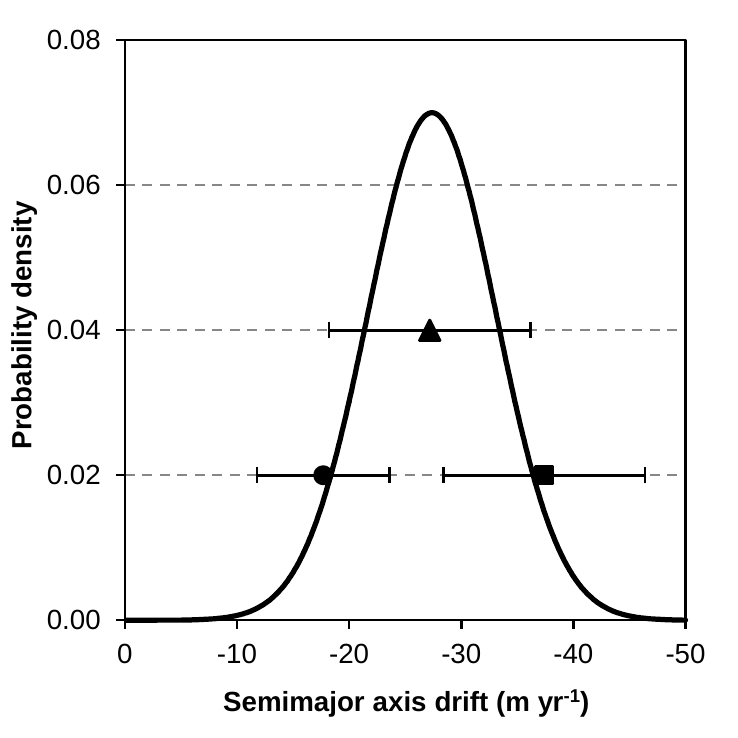}
\caption{Sampling distribution of the Yarkovsky semimajor axis drift (solid line) used in the analysis of (1620) Geographos compared against the three measured values and their uncertainties (circle - Chesley et al. 2008; square - Nugent et al. 2012; triangle - Farnocchia et al. 2013).}
\end{figure}

\begin{figure}
\centering
\includegraphics[width=\hsize]{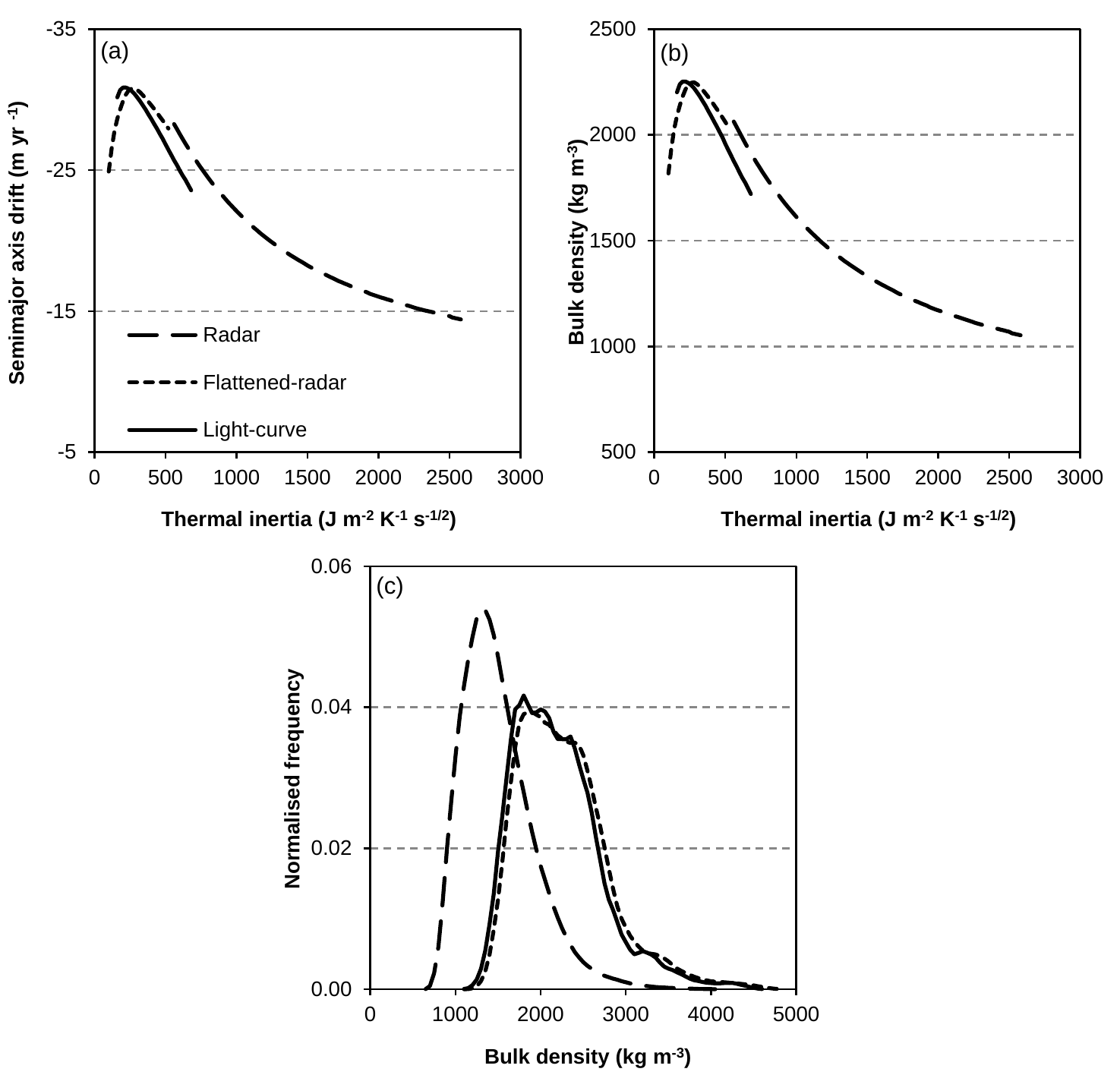}
\caption{Summary of ATPM Yarkovsky effect modelling results for (1620) Geographos using its three different shape models (legend) and a fixed maximum equatorial diameter of 5.04 $\pm$ 0.07 km. (a) Average Yarkovsky semimajor axis drift as a function of thermal inertia at a fixed bulk density of 2000 kg m$^{-3}$. (b) Average bulk density as a function of thermal inertia derived by comparing the model orbital drift against that measured. (c) The distribution of possible bulk densities derived at the 3-$\sigma$ confidence level.}
\end{figure}

\begin{figure}
\centering
\includegraphics[width=\hsize]{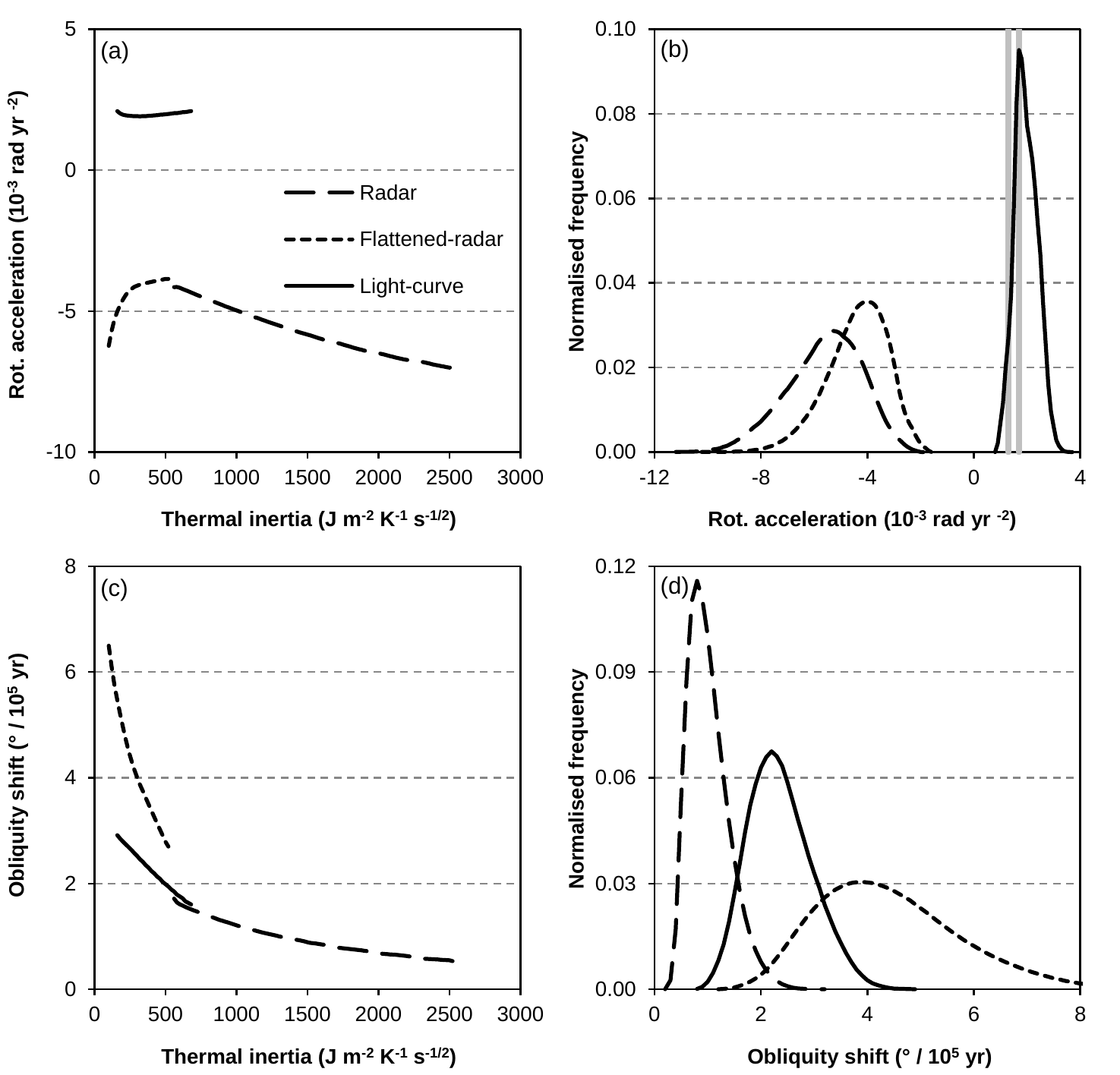}
\caption{Summary of ATPM YORP effect modelling results for (1620) Geographos using its three different shape models (legend) and a fixed maximum equatorial diameter of 5.04 $\pm$ 0.07 km. (a) Average YORP rotational acceleration as a function of thermal inertia. (b) The distribution of possible YORP rotational acceleration values derived at the 3-$\sigma$ confidence level. The grey vertical lines represent the lower and upper bounds of the YORP rotational acceleration acting on (1620) Geographos as measured by Ďurech et al. (2008b). (c) Average rate of YORP obliquity shift as a function of thermal inertia. (d) The distribution of possible rates of YORP obliquity shift derived at the 3-$\sigma$ confidence level.}
\end{figure}


\begin{thebibliography}{}

\bibitem{abe2006} Abe, S., Mukai, T., Hirata, N., et al. 2006, Science, 312, 1344
\bibitem{ali-lagoa2014} Alí-Lagoa, V., Lionni, L., Delbo, M., et al. 2014, A\&A, 561, A45
\bibitem{bottke1999} Bottke Jr., W. F., Richardson, D. C., Michel, P., \& Love, S. G. 1999, AJ, 117, 1921
\bibitem{bottke2006} Bottke Jr., W. F., Vokrouhlický, D., Rubincam, D. P., Nesvorný, D. 2006, Ann. Rev. Earth Planet. Sci., 34, 157
\bibitem{breiter2009} Breiter, S., Bartczak, P., Czekaj, M., et al. 2009, A\&A, 507, 1073
\bibitem{britt2002} Britt, D. T., Yeomans, D., Housen, K., Consolmagno, G. 2002, in Asteroids III, eds. W. F. Bottke Jr., A. Cellino, P. Paolicchi, \& R. P. Binzel (Tucson: Univ. of Arizona Press), 485
\bibitem{brozovic2010} Brozovic, M., Benner, L. A. M., Magri, C., et al. 2010, Icarus, 208, 207
\bibitem{bus2002} Bus, S. J., \& Binzel, R. P. 2002, Icarus, 158, 146
\bibitem{busch2007} Busch, M. W., Giorgini, J. D., Ostro, S. J., et al. 2007, Icarus, 190, 608
\bibitem{capek2004} Čapek, D., \& Vokrouhlický, D. 2004, Icarus, 172, 526
\bibitem{carry2012} Carry, B. 2012, P\&SS, 73, 98
\bibitem{chesley2014} Chesley, S. R., Farnocchia, D., Nolan, M. C., et al. 2014, Icarus, 235, 5
\bibitem{chesley2003} Chesley, S. R., Ostro, S. J., Vokrouhlický, D., et al. 2003, Science, 302, 1739
\bibitem{chesley2008} Chesley, S. R., Vokrouhlický, D., Ostro, S. J.,  et al. 2008, Asteroids, Comets, Meteors 2008, Baltimore, LPI Co. No. 1405, id. 8330
\bibitem{delbo2007} Delbo, M., Dell'Oro, A., Harris, A. W., et al. 2007, Icarus, 190, 236
\bibitem{durech2012b} Ďurech, J., Delbo, M., \& Carry, B. 2012b, Asteroids, Comets, Meteors 2012, Niigata, LPI Co. No. 1667, id. 6118
\bibitem{durech2008a} Ďurech, J., Vokrouhlický, D., Kaasalainen, M., et al. 2008a,  A\&A, 488, 345
\bibitem{durech2008b} Ďurech, J., Vokrouhlický, D., Kaasalainen, M., et al. 2008b, A\&A, 489, L25
\bibitem{durech2012a} Ďurech, J., Vokrouhlický, D., Baransky, A. R., et al. 2012a, A\&A, 547, A10
\bibitem{dunlap1974} Dunlap, J. L. 1974, AJ, 79, 324
\bibitem{emery2014} Emery, J. P., Fernández, Y. R., Kelley, M. S. P., et al. 2014, Icarus, 234, 17
\bibitem{farnocchi2013} Farnocchia, D., Chesley, S. R., Vokrouhlický, D., et al. 2013, Icarus, 224, 1
\bibitem{fowler1992} Fowler, J. W., \& Chillemi, J. R. 1992, in The IRAS Minor Planet Survey, eds. E. F. Tedesco, G. J. Veeder, J. W. Fowler, \& J. R. Chillemi (Hanscom AF Base: Philips Laboratory), 17
\bibitem{fujiwara2006} Fujiwara, A., Kawaguchi, J., Yeomans, D. K., et al. 2006, Science, 312, 1330
\bibitem{gaskell2008} Gaskell, R. 2008, Gaskell Itokawa Shape Model V1.0. HAY-A-AMICA-5-ITOKAWASHAPE-V 1.0, NASA Planetary Data System
\bibitem{green1985} Green, S. F. 1985, Ph.D. Thesis, University of Leicester
\bibitem{harris2005} Harris, A. W., Mueller, M., Delbo, M., \& Bus, S. J. 2005, Icarus, 179, 95 
\bibitem{hudson1999} Hudson, R. S., \& Ostro, S. J. 1999, Icarus, 140, 369
\bibitem{kaasalainen2007} Kaasalainen, M., Ďurech, J., Warner, B. D., et al. 2007, Nature, 446, 420
\bibitem{kaasalainen2013} Kaasalainen, M., \& Nortunen, H. 2013, A\&A, 558, A104
\bibitem{kaasalainen2001} Kaasalainen, M., \& Torppa, J. 2001, Icarus, 153, 24
\bibitem{kwiatkowski1995} Kwiatkowski, T. 1995, A\&A, 294, 274
\bibitem{lowry2012} Lowry, S. C., Duddy, S. R., Rozitis, B., et al. 2012, A\&A, 548, A12
\bibitem{lowry2007} Lowry, S. C., Fitzsimmons, A., Pravec, P., et al. 2007, Science, 316, 272
\bibitem{lowry2014} Lowry, S. C., Weissman, P. R., Duddy, S. R., et al. 2014, A\&A, 562, A48
\bibitem{magnusson1996} Magnusson, P., Dahlgren, M., Barucci, M. A., et al. 1996, Icarus, 123, 227
\bibitem{michalowski} Michalowski, T., Kwiatkowski, T., Borczyk, W., \& Pych, W. 1994, Acta Astronomica, 44, 223
\bibitem{muller2011} Müller, T. G., Ďurech, J., Hasegawa, S., et al. 2011, A\&A, 525, A145
\bibitem{muller2013} Müller, T. G., Miyata, T., Kiss, C. et al. 2013, A\&A, 558, A97
\bibitem{muller2005} Müller, T. G., Sekiguchi, T., Kaasalainen, M., et al. 2005, A\&A, 443, 347
\bibitem{muller2004} Müller, T. G., Sterzik, M. F., Schütz, O., Pravec, P., \& Siebenmorgen, R. 2004, A\&A, 424, 1075
\bibitem{nolan2013} Nolan, M. C., Magri, C., Howell, E. S., et al. 2013, Icarus, 226, 629
\bibitem{nugent2012} Nugent, C. R., Margot, J.-L., Chesley, S. R., \& Vokrouhlický, D. 2012, AJ, 144, 60
\bibitem{ostro2002} Ostro, S. J., Hudson, R. S., Benner, L. A. M., et al. 2002, in Asteroids III, eds. W. F. Bottke Jr., A. Cellino, P. Paolicchi, \& R. P. Binzel (Tucson: Univ. of Arizona Press), 151
\bibitem{ostro1996} Ostro, S. J., Jurgens, R. F., Rosema, K. D., et al. 1996, Icarus, 121, 46
\bibitem{ostro1995} Ostro, S. J., Rosema, K. D., Hudson, R. S., et al. 1995, Nature, 375, 474
\bibitem{rossi2009} Rossi, A., Marzari, F., \& Scheeres, D. J. 2009, Icarus, 202, 95
\bibitem{rozitis2013c} Rozitis, B., Duddy, S. R., Green, S. F., \& Lowry, S. C. 2013, A\&A, 555, A20
\bibitem{rozitis2011} Rozitis, B., \& Green, S. F. 2011, MNRAS, 415, 2042
\bibitem{rozitis2012} Rozitis, B., \& Green, S. F. 2012, MNRAS, 423, 367
\bibitem{rozitis2013a} Rozitis, B., \& Green, S. F. 2013a, MNRAS, 433, 603
\bibitem{rozitis2013b} Rozitis, B., \& Green, S. F. 2013b, MNRAS, 430, 1376
\bibitem{ryabova2002a} Ryabova, G. O. 2002a, Solar System Research, 36, 168
\bibitem{ryabova2002b} Ryabova, G. O. 2002b, Solar System Research, 36, 254
\bibitem{scheeres2008} Scheeres, D. J., \& Gaskell, R. W. 2008, Icarus, 198, 125
\bibitem{solem1996} Solem, J. C., \& Hills, J. G. 1996, AJ, 111, 1382
\bibitem{statler2009} Statler, T. S. 2009, Icarus, 202, 502
\bibitem{taylor2007} Taylor, P. A., Margot, J.-L., Vokrouhlický, D., et al. 2007, Science, 316, 274
\bibitem{tedesco2004} Tedesco, E. F., Noah, P. V., Noah, M., \& Price, S. D. 2004, IRAS Minor Planet Survey IRAS-A-FPA-3-RDR-IMPS-V6.0, NASA Planetary Data System 
\bibitem{veeder1989} Veeder, G. J., Hanner, M. S., Matson, D. L., Tedesco, E. F., et al. 1989, AJ, 97, 1211
\bibitem{vokrouhlicky2008} Vokrouhlický, D., Chesley, S. R., \& Matson, R.D. 2008, AJ, 135, 2336
\bibitem{vorder1993} Vorder Bruegge, R. W., \& Shoemaker, E. M. 1993, in AAS/Division for Planetary Sciences Meeting Abstracts, 25, \#23.17-P
\bibitem{wolters2011} Wolters, S. D., Rozitis, B., Duddy, S. R., et al. 2011, MNRAS, 418, 1246

\end{thebibliography}
\end{document}